\documentclass[a4paper,11pt]{article}
\usepackage{amsmath}
\usepackage{amsfonts}
\usepackage{amssymb}
\usepackage{bm}
\usepackage[utf8x]{inputenc}
\usepackage{ae}
\usepackage[T1]{fontenc}
\usepackage[english]{babel}
\usepackage{graphicx}
\usepackage{makecell}
\usepackage{placeins}
\usepackage{tabularx}
\usepackage{tabulary}
\usepackage{setspace}
\usepackage{tablefootnote}
\usepackage[flushmargin]{footmisc}
\usepackage[comma]{natbib}
\usepackage{array}
\usepackage{float}
\usepackage{caption}
\usepackage{tikz}
\usetikzlibrary{shapes.geometric}
\usepackage[a4paper,left=2.5cm,right=2.5cm,top=3cm,bottom=3cm]{geometry}
\usepackage[colorinlistoftodos]{todonotes}
\usepackage{pdfpages}
\usepackage{longtable}
\usepackage{adjustbox}
\usepackage{booktabs}
\usepackage{multirow}
\usepackage{lscape}
\usepackage[colorlinks=true, allcolors=blue]{hyperref}

\usepackage[flushleft]{threeparttable}
\usepackage{inputenc}



\begin{document} \doublespacing \pagestyle{plain}
	
	\def\ci{\perp\!\!\!\perp}
	\begin{center}
		
		{\LARGE The Impact of the UEFA Women’s EURO on Hotel Overnight Stays: Evidence from a Causal Analysis
		}

		{\large \vspace{0.8cm}}
		
		{\large Hannes Wallimann and Anna Mehr}\medskip

		{\small {University of Applied Sciences and Arts Lucerne, Institute for Tourism and Mobility} \bigskip }
		
		{\large \vspace{0.8cm}}
		
		{\small {September 2025} \bigskip }

	\end{center}
	
	\smallskip

	\small \noindent \textbf{Abstract:} {The impact evaluation of female sports events remains an important yet neglected area of research. To fill this gap, this working paper provides a timely assessment of the 2025 UEFA Women’s European Championship (WEURO) in Switzerland—the largest women-specific sports event in Europe with more than 657,000 spectators. Using city-level data on hotel overnight stays, we apply the Synthetic Difference-in-Differences approach of \citet{arkhangelsky2021synthetic} to compare WEURO host cities with non-host destinations. In summary, our results do not support strong claims of large tourism impacts but rather point to a small positive effect. Sensitivity analyses also suggest positive effects. However, confidence intervals permit firm conclusions only for the main venues, indicating an increase in overnight stays of 1.6\% attributable to the WEURO. Overall, our findings indicate positive but modest tourism impacts of the WEURO and outline a framework for further policy evaluation of sports events.}
	
	{\small \smallskip }
	{\small \smallskip }
	{\small \smallskip }
	
	
	{\small \smallskip }
	{\small \smallskip }
	{\small \smallskip }
	
	{\small \noindent \textbf{Acknowledgments:} The authors would like to thank Jürg Stettler for helpful comments.}
	
	\bigskip
	\bigskip
	\bigskip
	\bigskip
	
	{\small {\scriptsize 
\begin{spacing}{1.5}\noindent  
\textbf{Addresses for correspondence:} Hannes Wallimann, University of Applied Sciences and Arts Lucerne, Rösslimatte 48, 6002 Lucerne, \href{mailto:hannes.wallimann@hslu.ch}{hannes.wallimann@hslu.ch}. Anna Mehr, \href{mailto:anna.mehr@hslu.ch}{anna.mehr@hslu.ch}
\end{spacing}
			
		}\thispagestyle{empty}\pagebreak  }

	{\small \renewcommand{\thefootnote}{\arabic{footnote}} %
		\setcounter{footnote}{0}  \pagebreak \setcounter{footnote}{0} \pagebreak %
		\setcounter{page}{1} }
	
\section{Introduction}\label{introduction}

Local and national governments subsidize sports events with the justification that such events generate positive economic impacts and social benefits. Accordingly, policymakers are interested in how such events affect key performance indicators in tourism, such as tourism spending and overall economic growth. This also includes hotel-related indicators such as occupancy rates, hotel revenues, room prices, and restaurant activity \citep[see, e.g.,][]{dermody2003impact, lamla2014economic, depken2018hotel, falk2021short, collins2022impact}. However, the impact evaluation of female sports events remains an important but neglected area of research \citep[see, e.g.,][]{achu2022perceptions}. Exceptions include the studies by \citet{achu2022perceptions} and \citet{scott2011spectator}, which investigate the impacts of the 2016 Women’s Africa Cup of Nations hosted in Cameroon and the U.S. Women’s Open golf championship, respectively. 

We aim to fill this research gap by providing a timely impact evaluation of the 2025 UEFA Women’s European Championship (WEURO) in Switzerland---the largest women-specific sports event in Europe, with more than 657,000 spectators. To this end, we employ a the Synthetic Difference-in-Differences approach introduced by \citet{arkhangelsky2021synthetic}, to investigate the effect of a sports event on hotel overnight stays. To the best of our knowledge, this working paper is among the first to apply the Synthetic Difference-in-Differences approach to investigate the effect of sporting events on tourism indicators \citep[see also the recent study by][although that study does not examine a women’s sports event]{bradbury2025impact}. Using publicly available data on overnight stays from the Swiss Federal Statistical Office, we compare WEURO host venues with other Swiss destinations not affected by the event---an approach that, in contrast to survey-based impact evaluations, allows us to account for potential crowding-out effects. 

Investigating this aggregated macrodata of overnight stays, our results indicate that overnight stays in host cities increased by about 2,500 in July 2025, corresponding to a 1.9\% rise compared to the average overnight stays in the three years (July months) preceding the WEURO. The estimated effect is somewhat larger in absolute terms---around 3,400 additional overnight stays---and smaller in relative terms---1.6\% rise---among venues with bigger stadiums and a higher number of matches, namely Basel, Bern, Geneva, and Zurich. Sensitivity analyses, i.e., extended pre-intervention period and considering growth rates, also suggest positive effects. However, the 95\% confidence intervals are wide and include negative values when examining all venues, suggesting that the true effect cannot be determined with statistical precision. In contrast, the increase in overnight stays, when analyzing (only) main venues, is statistically significant at the 5\% level when applying a placebo inference procedure. In summary, the findings do not support strong claims of large tourism impacts, but rather suggest a small positive effect of the WEURO on overnight stays. 

The paper proceeds as follows. Section \ref{Literature} reviews the relevant literature. Section \ref{Background} discusses background information on the WEURO in Switzerland. Section \ref{Methods} outlines the methodological approach used to measure the effect on hotel overnight stays. The data employed for this purpose we present in Section \ref{Data}. Section \ref{Results} reports the findings. Finally, Section \ref{Discussion} discusses the results and concludes.

\section{Literature review}\label{Literature}

Major sporting events generate a wide range of measurable economic effects. Direct effects include revenue from ticket sales, merchandise sales, sponsorship deals and tourist spending. Indirect effects stem from the creation of local economic value, employment opportunities, and infrastructure investment. Meanwhile, induced effects reflect the long-term impact on tourism, destination image, and regional purchasing power. It is important for policymakers to analyze these effects, which is why many studies have been conducted on this topic. Closely related to our study are papers that use real-world data to estimate the impact on the hotel market.

The first paper using daily hotel occupancy data appears to be the study of \citet{dermody2003impact}, which analyzes the impact of all NFL teams in their home cities. They conclude that most NFL teams positively affect hotel occupancy and revenue. Moreover, \citet{depken2018hotel} show that two political conventions and NASCAR auto races in Charlotte, North Carolina are associated with large increases in hotel occupancy, prices, and revenue, whereas other events have no effect on the hotel market. More recently, \citet{collins2022impact} use daily hotel occupancy data to investigate the impact of sports and cultural events in Austin, Texas. The results indicate disparities between events with respect to the quality of hotels demanded. For instance, football games hosted at the University of Texas are associated with a statistically significant increase in the number of rooms rented on game days, with the largest effect for Upper Middle Tier hotels. \citet{falk2021short} examine the effect of sports events on hotel room prices in Finnish Lapland. Using the Difference-in-Differences method, they show that, on average, hotel room prices rise by 14\% during the event. Their results also indicate effect heterogeneity: small-scale events show no impact, while the strongest effect is found for the Levi FIS Alpine Ski World Cup competition. In addition, \citet{deschriver2021sporting} demonstrate that the impact of Southeastern Conference collegiate football games on hotel performance varies greatly according to team and opponent quality.

Closely related to our study is the paper by \citet{lamla2014economic}, which investigates the economic impact of UEFA European Championship 2008 in Switzerland (the event also took place in Austria, but the study focuses on Switzerland). Using firm-level data from more than 700 Swiss hotels and restaurants, \citet{lamla2014economic} find negative but heterogeneous effects. Moreover, descriptive analyses of aggregated macrodata do not reveal any economic impact. They conclude that the disappointing results for the hotel sector may be related to the fact that many congresses and seminars that would usually have been held in June were rescheduled or cancelled because of the tournament. In addition, the event may have induced crowding-out effects in the hotel sector.

However, analyses of major female sports events are far less common---according to \citet{achu2022perceptions}, this represents an important but neglected area of research. One exception is the study by \citet{scott2011spectator}, which investigates the economic impact (and spectator profiles) of the 2009 U.S. Women’s Open. They report spectator spending of USD 7.6 million, with overnight visitors accounting for USD 4.3 million of this amount. Another example is the study by \citet{achu2022perceptions}, which examines the socioeconomic impact of the 2016 Women’s Africa Cup of Nations hosted in Cameroon. According to their findings, the event increased short-run employment and income opportunities in the host cities. However, to the best of our knowledge, the causal economic effects of sports events exclusively involving women have not yet been discussed in the literature.

Note that the economic benefits of major events are often overstated due to methodological issues, such as ignoring displacement and crowding-out effects, applying incorrect multipliers, and failing to account for opportunity costs (see, e.g., \citet{crompton1995economic} or \citet{preuss2005economic}). A systematic review by \citet{zourgani2023systematic} likewise shows that many studies exaggerate the economic impact and rarely consider long-term consequences.

More broadly, our study is also related to the literature that investigates the effects of sports events using the synthetic control method introduced by \citet{abadie2010synthetic}. For instance, \citet{kobierecki2022sports}, applying the generalized synthetic control method of \citet{xu2017generalized}, conclude that staging the Olympic Games (London, Rio de Janeiro, and Vancouver) or the FIFA World Cup (Brazil and South Africa) between 2010 and 2016 had no significant effect on economic growth. Moreover, \citet{breidenbach2022large} apply the synthetic control method in a sensitivity analysis to assess the effect of large-scale sports events on COVID-19 infections. Another example is the study by \citet{zhang2022impacts}, which analyzes the impact of sports events on local carbon emissions using the case of the Nanjing Youth Olympics. Finally, \citet{bradbury2025impact} analyze the relocation of an Major League Baseball team and its impact on hotel demand and concludes that no robust increases in overnight stays can be observed at the regional level. The study thus provides, to the best of our knowledge, a unique example of applying the recent Synthetic Difference-in-Differences approach of \citet{arkhangelsky2021synthetic} to tourism outcomes.

\section{Background}\label{Background}

The UEFA Women’s European Championship (WEURO) took place in July 2025, featuring a total of 31 matches (see Table \ref{tab:matches}). Overall, it was the highest-attended Women’s EURO in history, attracting 657,291 spectators to the stadiums. This record surpassed the previous benchmark of 574,875 fans at the 2022 tournament in England and the 240,055 fans at the 2017 edition in the Netherlands.\footnote{See \href{https://www.uefa.com/womenseuro/news/029b-1e522ecda3bd-ccdd22a6bb3c-1000--milestones-met-history-made-record-breaking-women-s-euro-rea/}{UEFA (2025): Milestones met, history made – record-breaking Women’s EURO}, accessed on August~28,~2025.} 
On average, about 21,000 spectators attended each match. The final, held in Basel---where England defeated Spain in a penalty shoot-out---drew a crowd of 34,203.

Basel also hosted the largest stadium, with a capacity of 34,250 seats (see Table \ref{tab:matches}), followed by Bern (29,800), Geneva (26,750), and Zurich (22,700). These main venues staged the majority of matches, including the opening match in Basel on 2 July and the final on 27 July. The two semifinals were held in Geneva and Zurich on 22 and 23 July.

Overall, the UEFA Women's EURO took place across eight stadiums in Switzerland. Additional host cities included St. Gallen, Lucerne, Thun, and Sion, where smaller stadiums with capacities ranging from 16,300 to 7,750 seats were used exclusively for group-stage matches.\footnote{Note that all stadiums already existed. Therefore, the WEURO, in contrast to other major sporting events, did not trigger any significant investments in Swiss infrastructure. Only minor adjustments were made---for example, the stadium in Bern received a real grass pitch, which was arguably necessary for important football matches.} In addition to the spectators inside the stadiums, the host venues also organized public viewing areas and official fan zones.  

According to Visa navigate, the  travellers to Switzerland rose by 12\% during the opening week of the UEFA Women’s EURO.\footnote{See \href{https://navigate.visa.com/europe/research-and-insights/uefa-womens-euro-2025-the-economic-ripple-effect-from-stadium-to-storefront/?utm_source=chatgpt.com}{UEFA Women's Euro 2025: The Economic Ripple Effect from Stadiums to Storefront}, accessed on September~23,~2025.} Moreover, Visa cardholders from countries participating in UEFA Women’s EURO 2025 spent more on average year-on-year in Switzerland during the first week of the tournament with the largest increase of overseas spend comes from Iceland with a nearly fivefold year-on-year increase. 

Moreover, a report by EY of UEFA Women's EURO 2025 shows added value of CHF 205 million for Switzerland (forecast: CHF 180 million).\footnote{See, e.g.,  \href{https://editorial.uefa.com/resources/0299-1dbce1a4fb6f-5384a0ce3fd8-1000/uefa_weuro_2025_approaching_the_summit_en_may_2025_.pdf}{Pre-tournament impact report}, accessed on September~25,~2025.} 35\% of tickets were sold to international guests.\footnote{See, e.g., \href{https://weuro2025summary.uefa.com}{UEFA Women's EURO Tournament Suammry}, accessed on September~25,~2025.} According to UEFA data, Swiss hotels recorded a 9\% increase in European bookings during the tournament month compared to July 2024.

\begin{table}[htbp]
	\centering
	\caption{UEFA Women’s EURO 2025 — Matches by venue}\label{tab:matches}
	\tiny
	\setlength{\tabcolsep}{2pt}
	\renewcommand{\arraystretch}{1.05}
	\begin{tabular}{@{}p{0.32\textwidth}p{0.32\textwidth}p{0.32\textwidth}@{}}
		\toprule
		\textbf{St. Jakob-Park (Basel, cap. 34\,250)} &
		\textbf{Stadion Wankdorf (Bern, cap. 29\,800)} &
		\textbf{Stade de Genève (Geneva, cap. 26\,750)} \\
		\midrule
		\begin{tabular}[t]{@{}l@{}}
			2.7: SUI–NOR (1:2)\\
			8.7: GER–DEN (2:1)\\
			13.7: NED–FRA (2:5)\\
			19.7: QF: FRA–GER (1:1, 5:6 pens)\\
			27.7: Final: ENG–ESP (1:1, 3:1 pens)
		\end{tabular}
		&
		\begin{tabular}[t]{@{}l@{}}
			3.7: ESP–POR (5:0)\\
			6.7: SUI–ISL (2:0)\\
			11.7: ITA–ESP (1:3)\\
			18.7: QF: ESP–SUI (2:0)
		\end{tabular}
		&
		\begin{tabular}[t]{@{}l@{}}
			4.7: DEN–SWE (0:1)\\
			7.7: POR–ITA (1:1)\\
			10.7: FIN–SUI (1:1)\\
			16.7: QF: NOR–ITA (1:2)\\
			22.7: SF: ENG–ITA (2:1 a.e.t.)
		\end{tabular} \\
		\addlinespace[4pt]
		\textbf{Letzigrund (Zurich, cap. 22\,700)} &
		\textbf{Arena St.Gallen (St. Gallen, cap. 16\,300)} &
		\textbf{Allmend Stadion Luzern (Lucerne, cap. 14\,350)} \\
		\midrule
		\begin{tabular}[t]{@{}l@{}}
			5.7: FRA–ENG (2:1)\\
			9.7: ENG–NED (4:0)\\
			12.7: SWE–GER (4:1)\\
			17.7: QF: SWE–ENG (2:2, 2:3 pens)\\
			23.7: SF: GER–ESP (0:1 a.e.t.)
		\end{tabular}
		&
		\begin{tabular}[t]{@{}l@{}}
			4.7: GER–POL (2:0)\\
			9.7: FRA–WAL (4:1)\\
			13.7: ENG–WAL (6:1)
		\end{tabular}
		&
		\begin{tabular}[t]{@{}l@{}}
			5.7: WAL–NED (0:3)\\
			8.7: POL–SWE (0:3)\\
			12.7: POL–DEN (3:2)
		\end{tabular} \\
		\addlinespace[4pt]
		\textbf{Arena Thun (Thun, cap. 8\,100)} &
		\textbf{Stade de Tourbillon (Sion, cap. 7\,750)} &
		\multicolumn{1}{c}{} \\
		\midrule
		\begin{tabular}[t]{@{}l@{}}
			2.7: ISL–FIN (0:1)\\
			7.7: ESP–BEL (6:2)\\
			10.7: NOR–ISL (4:3)
		\end{tabular}
		&
		\begin{tabular}[t]{@{}l@{}}
			3.7: BEL–ITA (0:1)\\
			6.7: NOR–FIN (2:1)\\
			11.7: POR–BEL (1:2)
		\end{tabular}
		&
		\multicolumn{1}{c}{} \\
		\bottomrule
	\end{tabular}
	\begin{tablenotes}[flushleft]
		\footnotesize
		\item \textbf{Notes:} Source: \url{https://de.uefa.com/womenseuro}, assessed on August 22, 2025. 'Cap.' denotes the capacity of the stadium. 
	\end{tablenotes}
\end{table}

Sustainability was another key focus of the WEURO. To support this goal, the Swiss Federal Office of Transport provided a subsidy of CHF~5~million to the public transport industry.\footnote{See \url{https://www.baspo.admin.ch/de/newnsb/GekvZn8I7W9PQMcGWdKiY}, accessed on August 28, 2025.} This funding enabled ticket holders to travel free of charge from anywhere in Switzerland to the stadium and back on match days. The offer was valid across the entire Swiss public transport network. This so-called KombiTicket initiative proved highly successful, with about 66\% of ticket holders travelling to the stadiums by public transport.\footnote{See, e.g., \href{https://www.uefa.com/womenseuro/news/029b-1e522ecda3bd-ccdd22a6bb3c-1000--milestones-met-history-made-record-breaking-women-s-euro-rea/}{UEFA Women’s EURO milestones}, accessed on August 28, 2025.} In addition, around 20\% of spectators reached the stadiums on foot or by bicycle.  

In total, the Swiss federal government supported the WEURO with CHF~15~million.\footnote{See \url{https://www.baspo.admin.ch/de/uefa-womens-euro-2025-in-der-schweiz-die-rolle-des-bundes}, accessed on September 1, 2025.} Of this amount, in addition to the CHF~5 million for the public transport industry, CHF~10~million were allocated to Switzerland Tourism for international promotional activities and to the Swiss Football Association for the further development of women’s football. Moreover, Swiss cities and cantons provided a total of CHF~63 million to the event.

\section{Identification and estimation}\label{Methods}

In our study, we aim to estimate the effect of the binary treatment "being a WEURO venue" on overnight stays in these venues in July 2025, in order to assess the economic impact of the sports event. Using statistical parlance, we are interested in the effect on the subpopulation of venues that received the treatment, i.e., the WEURO, corresponding to the so-called Average Treatment Effect on the Treated ($ATT$). Formally, the $ATT$ can be expressed as

\[
\Delta_{D=1} = E[Y(1) - Y(0) \mid D = 1],
\]

where $Y(1)$ and $Y(0)$ denote the potential outcomes (overnight stays) under treatment and no treatment, respectively. The indicator $D=1$ signifies that the effect is estimated among venues that actually hosted the WEURO. While we can observe in the data the outcome under treatment, i.e., the overnight stays of the venues in 2025, we must estimate the so-called counterfactual outcome: the overnight stays that would have occurred at these venues had the WEURO not taken place.

To do so, we apply the Synthetic Difference-in-Differences method proposed by \citet{arkhangelsky2021synthetic}. Put simply, this method, in a data-driven way, selects which version (or combination) of the synthetic control approach \citep[see, e.g.,][]{abadie2010synthetic} performs best, using a set of control venues not exposed to the treatment ($D=0$) to mimic the evolution of the treated venues’ overnight stays \citep[for an intuitive explanation see, e.g., ][]{huber2025impact}.  

One version is the Synthetic Control Method proposed by \citet{abadie2010synthetic}. This method constructs a synthetic venue by assigning weights to control venues in order to best approximate $Y(0)$, the potential overnight stays of the WEURO venues in the absence of the event. The weights are chosen such that the treated and control groups are as similar as possible in the pre-intervention period. Applying the synthetic control method requires the assumption that the characteristics of the treated and control venues are comparable (so-called selection-on-observables assumption). In addition, the overnight stays of the WEURO venues must not be too extreme compared to those of the control venues (also called the convex hull condition). (Note that the Synthetic Control Method is often applied in settings with a single treated unit. However, the underlying idea in our case remains the same.)

On the other hand, the Synthetic Difference-in-Differences method may select an alternative that is closer to the Difference-in-Differences approach \citep[see, e.g., ][]{ashenfelter1978estimating}. In this case, the potential outcome of the venues under no treatment, $Y(0)$, is constructed such that the trends in the pre-intervention periods of control and treated venues follow a common path. To assess the impact of the WEURO, we then compare the before–after changes in overnight stays between the treated venues and control venues in the control group. This approach relies on the so-called common trend assumption, while allowing the levels of overnight stays between treatment and control venues to differ.  

To estimate the causal effect of the WEURO on overnight stays, using the Synthetic Difference-in-Differences method, we apply the \textsf{synthdid} package in the statistical software \textsf{R}.

\section{Data and descriptive statistics}\label{Data}

To assess the impact of the 2025 UEFA Women's European Championship (WEURO) on overnight stays, we use municipality-level monthly panel data on overnight stays provided by the Swiss Federal Statistical Office (FSO).\footnote{See \url{https://www.bfs.admin.ch/bfs/de/home/statistiken/katalog.assetdetail.36162296.html}, accessed on September 9, 2025.} We focus on the month of July for the 100 destinations with the highest number of overnight stays in 2024. From this set of municipalities, we select the (relatively) urbanized municipalities (henceforth called cities), which constitute the treatment and control groups. The treatment group consists of Zurich, Basel, Bern, Geneva, Lucerne, St. Gallen, Thun, and Sion (see also Table \ref{tab:matches}). The control group includes Lausanne, Winterthur, Neuchâtel, Chur, Biel/Bienne, Solothurn, Vevey, Schaffhausen, Baden, Bellinzona, Nyon, Meyrin, Vernier, Rümlang, Wallisellen, Bussigny, and Lugano. Note that due to the COVID-19 pandemic, we restrict our pre-intervention period from 2022 to 2024.

In our case, we observe that some of the cities with the most overnight stays are venues affected by the treatment, and are more extreme---compared to the destinations in the control group---in the values of the overnight stays before the Women's EURO (see Table \ref{tab:summary_stats}). Therefore, there will not be a weighted average of untreated units that can approximate the overnight stays for the treated units before the intervention. Thus, the convex hull condition would be violated. Therefore, following \citet{abadie2021using}, \citet{ferman2019inference}, and \citet{wallimann2022complementary} (in the tourism literature), we measure overnight stays in differences with respect to pre-intervention means, i.e., 

\[
\tilde{Y}_{jt} \;=\; Y_{jt} \;-\; \frac{1}{T_0} \sum_{i=1}^{T_0} Y_{ji}, 
\]

where $Y_{jt}$ denotes the observed value of city $j$ at time $t$. The term $\frac{1}{T_{0}} \sum_{i=1}^{T_{0}} Y_{ji}$ represents the average outcome of unit $j$ during the $T_{0}$ pre-intervention periods. Accordingly, $\tilde{Y}_{jt}$ measures the overnight stays of city $j$ at time $t$ expressed as a deviation from its pre-intervention mean. This allows the Synthetic Difference-in-Differences approach also to rely on the 'classic' Synthetic Control Method by \citet{abadie2010synthetic}.\footnote{Note that in a sensitivity analysis, we re-estimate the impact of the WEURO using growth rates of overnight stays, defined as $g_{jt} = \tfrac{Y_{jt}}{Y_{j,\,t-1}} - 1$, where $g_{jt}$ denotes the year-on-year growth rate of unit $j$ at time $t$.}\label{footnote:growth} Table \ref{tab:summary_stats} reports mean overnight stays and their standard deviations for treated and control cities in the pre- and post-intervention periods. While both groups show an increase in average overnight stays in 2025, the rise is substantially larger for the treated host cities.

\begin{table}[htbp]
	\centering
	\begin{threeparttable}
		\caption{Summary statistics of overnight stays by treatment status and period}
		\label{tab:summary_stats}
		\begin{tabular}{lcccccc}
			\toprule
			Group & Period & Mean ONS & SD ONS & Mean ONS diff & SD ONS diff & n \\
			\midrule
			Control            & Pre  & 24,262  & 25,989  & 0     & 3,894  & 57 \\
			Control            & Post & 26,594  & 28,522  & 2,332 & 3,700  & 19 \\
			Treated (all hosts)& Pre  & 134,249 & 122,598 & 0     & 11,101 & 24 \\
			Treated (all hosts)& Post & 144,444 & 134,596 & 10,195 & 9,321 & 8 \\
			Treated (main hosts)& Pre  & 216,693 & 113,985 & 0     & 15,409 & 12 \\
			Treated (main hosts)& Post & 232,553 & 126,205 & 15,860 & 6,060 & 4 \\
			\bottomrule
		\end{tabular}
		\begin{tablenotes}
			\footnotesize
			\item \textit{Notes}: SD = standard deviation. ONS = overnight stays. ONS diff refers to overnight stays expressed as deviations from pre-intervention means. n is the number of municipality-year observations.
		\end{tablenotes}
	\end{threeparttable}
\end{table}

Note that determining the statistical significance of Synthetic Control Method-based estimates is not straightforward \citep[see, e.g., ][]{huber2023causal}. In our paper, we apply a placebo inference procedure, where we consider the cities in the control group to be pseudo-treated.

\section{Results}\label{Results}

\subsection{Impact of the WEURO across all host cities}

\begin{figure}[htbp]
	\centering
	\includegraphics[width=0.8\textwidth]{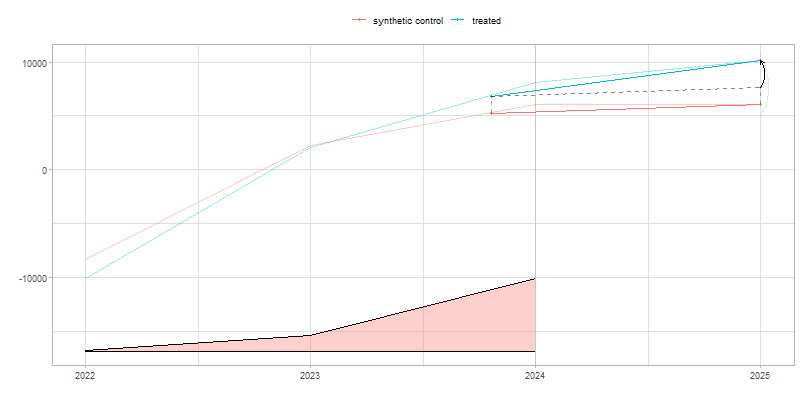}
	\caption{Impact of the WEURO on overnight stays in July 2025}
	\label{fig:sdid:mean}
\end{figure}

In Figure \ref{fig:sdid:mean}, showing the results when applying the Synthetic Difference-in-Differences method, we see the trajectories of all host cities (blue, treated) and their synthetic counterfactual constructed from the cities in the control group (red), with 2025 representing the intervention period. The relative weights assigned to the pre-intervention periods are displayed in the lower panel of Figure \ref{fig:sdid:mean}, showing that the year 2024 received the highest weight of 0.813. The diagram of the estimate is overlaid: the blue segment depicts the change from the weighted pre-intervention average to the post-intervention average of the host cities, while the red segment shows the corresponding change for the control cities. The counterfactual trajectory---represented by the dashed segment parallel to the red line in Figure \ref{fig:sdid:mean}---is derived from the evolution of the control cities. Comparing this counterfactual with the actual observed overnight stays of the treated cities in July 2025 yields an average treatment effect on the treated (ATT) of 2,536 additional overnight stays. This represents a 1.9\% increase in overnight stays compared to the pre-treatment number of overnight stays (see Table \ref{tab:summary_stats}). 

However, we obtain a standard error of 1,388, which is quite substantial relative to the absolute magnitude of the causal effect. Accordingly, the 95\% confidence interval ranging from -185 to 5257---represented by the grey arrows in Figure \ref{fig:sdid:mean}---includes negative values, suggesting that the estimated positive effect on overnight stays may be driven by chance. However, the impact of the WEURO is statistically significant at the 10\% level with a confidence interval ranging from 252 to 4,820.

\subsection{Impact of the WEURO in main host cities}

It is possible that the effect was substantially larger in cities with bigger stadiums and a higher number of matches. At the WEURO, these were Basel, Bern, Geneva, and Zurich, each hosting at least four matches, including one in the knockout stage. Therefore, we re-estimate the impact of the WEURO for these venues only. (Note that, for this estimation, we omit Lucerne, St. Gallen, Thun, and Sion.)

\begin{figure}[htbp]
	\centering
	\includegraphics[width=0.8\textwidth]{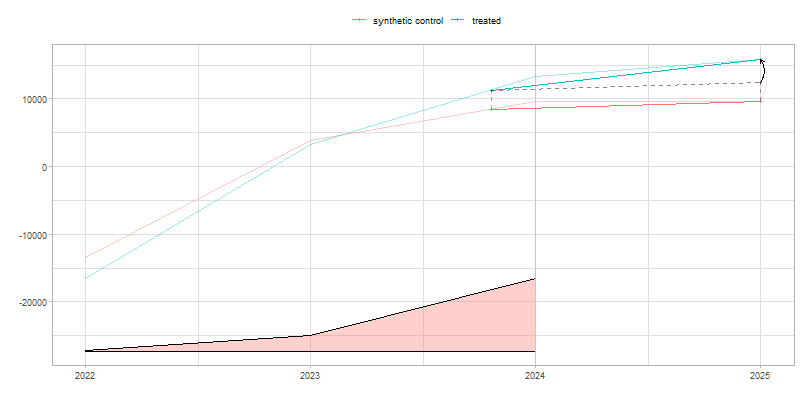}
	\caption{Impact of the WEURO on overnight stays in July 2025 (main host cities)}
	\label{fig:sdid:mean:main}
\end{figure}

The estimated impact of the WEURO now amounts to 3,419 additional overnight stays, with 2024 again receiving the highest time weight (see Figure~\ref{fig:sdid:mean:main}). This corresponds to an increase of 1.6\% relative to the average number of overnight stays of the main hosts in the pre-intervention period (2022--2024) (see Table \ref{tab:summary_stats}). The 95\% confidence interval ranges from 843 to 6830. Accordingly, we now conclude that the effect is statistically significant.

\subsection{Sensitivity analysis}\label{results:sensitivity}

Due to the COVID-19 pandemic, we restricted our pre-intervention period to the years 2022--2024. In a first sensitivity analysis, we expand the pre-intervention period to include the years 2017--2024.\footnote{Note that we had to remove Wallisellen and Rümlang, as their time series are not complete.} The estimated impact on overnight stays (for all host cities) remains positive but is smaller, amounting to 333 additional overnight stays. The effect is now far from statistically significant, with 95\% confidence intervals ranging from $-2{,}205$ to $2{,}871$. Figure~\ref{fig:sdid:mean:expanded} in Appendix~\ref{Appendix_Figures} illustrates this effect. It also clearly shows that the decline in overnight stays during the COVID-19 period was considerably more pronounced among the host cities compared to the control cities, suggesting that our main study design with a shorter pre-intervention period is appropriate.  

In a second sensitivity analysis, we replace $\tilde{Y}_{jt}$, the measure of overnight stays expressed as deviations from pre-intervention means, with the growth rates $g_{jt}$ (see Footnote~\ref{footnote:growth}). This specification suggests an additional growth in overnight stays of 4.8\% (see Figure~\ref{fig:sdid:mean:growth} in Appendix~\ref{Appendix_Figures}), which is even higher than in our main results. However, the 95\% confidence interval ranges from $-0.16$ to $0.26$ and therefore includes negative values.

\section{Discussion and Conclusion}\label{Discussion}

In this paper, we applied the synthetic difference-in-differences approach to investigate the impact of the 2025 UEFA Women’s European Championship (WEURO) on host cities in Switzerland. Our analysis indicates that overnight stays increased by about 2,500 in July, which is still meaningful, albeit corresponding to only a 1.9\% rise compared to pre-intervention levels. The positive effect is even larger in absolute terms—around 3,400 additional overnight stays—among venues with bigger stadiums and a higher number of matches, namely Basel, Bern, Geneva, and Zurich. In relative terms, however, this corresponds to only a 1.6\% increase in overnight stays. Sensitivity analyses also suggest positive results. However, when calculating 95\% confidence intervals using a placebo inference procedure, we find that the uncertainty is substantial and does not allow firm statistical conclusions about whether the true effect is positive for all venues. In contrast, the impact on the main venues appears statistically significant at the 5\% level. In summary, our results do not support strong claims of large tourism impacts but instead point to a small positive effect, particularly for destinations with bigger stadiums and more matches.

However, a small effect does not necessarily imply that the WEURO had no impact. On the one hand, it is possible that spectators (e.g., the approximately 230,000 from abroad, see Section \ref{Background}) did not stay in hotels but rather, for example, in holiday flats. However, this would imply that the economic effects within the destination’s value network were smaller. On the other hand, because of the free public transport offer, spectators were able to travel at lower costs. As a result, overnight stays may not have occurred directly at the match venues, and so-called spillover effects cannot be fully excluded. Moreover, there could still have been economic effects, for instance, if hotel prices increased, as shown by \citet{falk2021short}, which should be investigated in future research.  

Another explanation could be (unintended) crowding-out effects: It may be possible that the hotels in the host venues would have been fully booked (or accommodated many guests) even without the WEURO. Then, the visitors drawn by a sports event may displace others who would have visited but did not because they could not score accommodations or they were not willing to deal with the crowds of the sports event. Studies of major sports events show that high overnight occupancy rates do not necessarily indicate genuine additional demand. For example, \citet{scott2011spectator} identified a slight crowding-out effect during the 2009 US Women's Open golf championship. \citet{fourie2011tourist} also found that visitors from non-participating countries avoided two South African events, while fans from participating countries simply postponed their trips. Using the example of the World Cup, \citet{preuss2011method} points out that crowding out in the narrow sense only affects those who actually cancel their trip, and distinguishes this from time and location shifts. He works with clearly defined visitor groups ('football visitors', 'extenders', 'cancellers', 'pre/post switchers', etc.) and combines surveys with official overnight statistics to determine net effects. For the Women's EURO in Switzerland, this means that high hotel occupancy rates could be due to shifts in time or location among other guest segments. Therefore, future research should separate these effects to assess the economic value added by the tournament.

A limitation of our study is that, due to data restrictions, we used aggregated monthly hotel data. Daily data would allow for the incorporation of variables for the days leading up to or following an event, i.e., a football game, and to better capture the duration of visitors’ stays \citep{collins2022impact}. Therefore, future research should investigate the effect of the UEFA Women's EURO---or another female sports event---using daily data. 

Future research should also aim to causally analyze whether the free arrival and return travel to the matches had an effect on transport mode choice. Such research would be closely linked to transportation studies that examine mobility offers to and within tourism destinations. An example is the paper by \citet{blattler2024free}, which analyzes how a free arrival and departure offer of a Swiss tourism destination affects transport mode choice. One could also interpret this type of offer more broadly as a guest card initiative for tourists at a destination \citep{gronau2017encouraging}. Further research could also examine whether free arrival and return travel influences overnight stays. For instance, does such an offer encourage visitors from farther away to travel to and from the destination on the same day (day trips), thereby reducing the number of overnight guests?

This study focuses on short-run economic effects. Future research should also investigate long-run effects. Moreover, sports events not only focus on economic goals but also on a wide array of goals, such as social, political, cultural, and environmental objectives, for host communities \citep{zourgani2023systematic}. Therefore, the other non-economic impacts of the WEURO should also be examined, for example, with respect to gender equality in sports.  

On the other hand, our study design could be applied in future research to investigate and compare the (short and long-run) impacts of different events within Switzerland and to examine differences between them. Moreover, due to the large number of major events taking place in Switzerland, including the 2026 Ice Hockey World Championship and the Mountain Bike World Championship in Valais, it would be beneficial to replicate the analytical concept developed in future studies. A comparative analysis of several events would enable the systematic identification of patterns and differences in tourist demand.

Finally, we note that we focused on overnight stays from both domestic and international tourists, as our aim was to measure the overall economic impact. Future studies could, however, concentrate exclusively on international visitors, and an analysis of arrivals may also be of interest.

	\newpage
	\bigskip
	
	\bibliographystyle{sageh}
	\bibliography{WEURO.bib}
	
	\bigskip
	\newpage
	
\begin{appendix}
		
		\numberwithin{equation}{section}
		\counterwithin{figure}{section}
		\noindent \textbf{\LARGE Appendices}

	\section{Additional Figures}\label{Appendix_Figures}

\begin{figure}[htbp]
	\centering
	\includegraphics[width=0.8\textwidth]{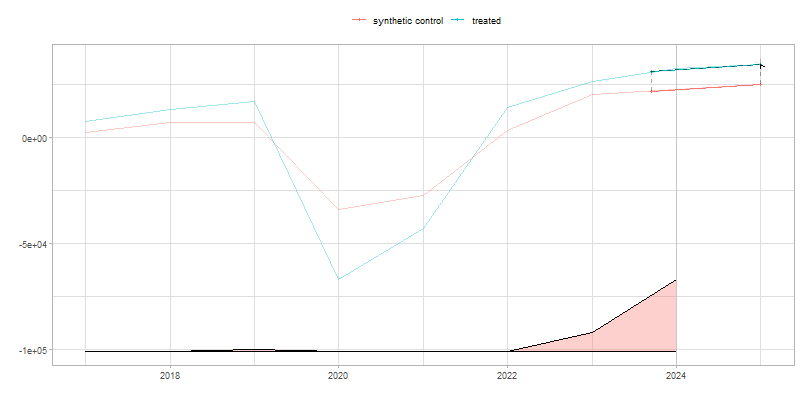}
	\caption{Impact of the WEURO on overnight stays in July 2025 (expanded pre-intervention period)}
	\label{fig:sdid:mean:expanded}
\end{figure}

\begin{figure}[htbp]
	\centering
	\includegraphics[width=0.8\textwidth]{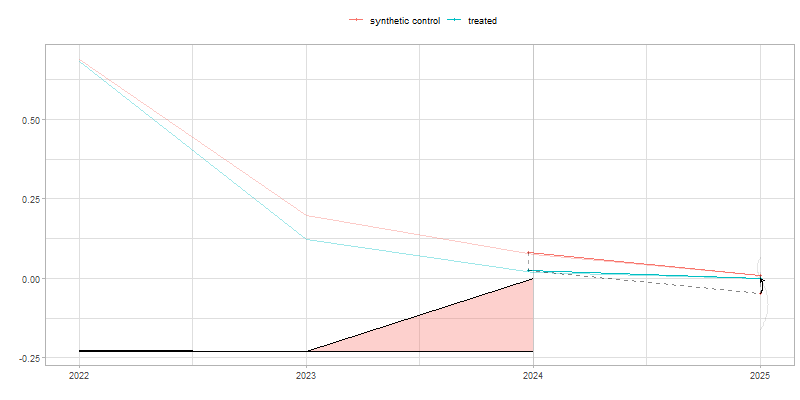}
	\caption{Impact of the WEURO on overnight stays in July 2025 (growth rates)}
	\label{fig:sdid:mean:growth}
\end{figure}

	\end{appendix}
\end{document}